\documentclass[12pt,draftcls,peerreviewca,onecolumn,a4paper,dvips]{IEEEtran}
\usepackage{stmaryrd}
\usepackage{amsfonts}
\usepackage{amssymb}
\usepackage{stfloats}
\usepackage{cite}
\usepackage{graphicx}
\usepackage{psfrag}
\usepackage{subfigure}
\usepackage{amsmath}
\usepackage{array}
\usepackage{eurosym}
\begin{document}

\title{Fundamental Tradeoffs on Green Wireless Networks}

\newtheorem{Thm}{Theorem}
\newtheorem{Lem}{Lemma}
\newtheorem{Cor}{Corollary}
\newtheorem{Def}{Definition}
\newtheorem{Exam}{Example}
\newtheorem{Alg}{Algorithm}
\newtheorem{Prob}{Problem}
\newtheorem{Rem}{Remark}

\author{{Yan Chen, Shunqing Zhang, Shugong Xu, and Geoffrey Ye Li}
\thanks{Yan Chen, Shunqing Zhang and Shugong Xu (e-mail: {\tt \{eeyanchen, sqzhang, shugong\}@huawei.com}) are with the GREAT (Green Radio Excellence in Architectures and Technologies) Research Team of Huawei Technologies Co., Ltd., Shanghai, China; Geoffrey Ye Li (email: {\tt {liye@ece.gatech.edu}}) is with the School of ECE, Georgia Institute of Technology, Atlanta, GA, USA.}}

\maketitle

\begin{abstract}
Traditional design of mobile wireless networks mainly focuses on ubiquitous access and large capacity. However, as energy saving and environmental protection become a global demand and inevitable trend, wireless researchers and engineers need to shift their focus to energy-efficiency oriented design, that is, green radio. In this paper, we propose a framework for green radio research and integrate the fundamental issues that are currently scattered. The skeleton of the framework consists of four fundamental tradeoffs: deployment efficiency - energy efficiency tradeoff, spectrum efficiency - energy efficiency tradeoff, bandwidth - power tradeoff, and delay - power tradeoff. With the help of the four fundamental tradeoffs, we demonstrate that key network performance/cost indicators are all stringed together.
\end{abstract}


\newpage

\section{Introduction}\label{sec:intro}

\subsection{Why Green Evolution?}
The next generation wireless networks are expected to provide high speed internet access anywhere and anytime. The popularity of iPhone and other types of smartphones doubtlessly accelerates the process and creates new traffic demand, such as mobile video and gaming. The exponentially growing data traffic and the requirement of ubiquitous access have triggered dramatic expansion of network infrastructures and fast escalation of energy demand. Hence, it becomes an urgent need for mobile operators to maintain sustainable capacity growth and, at the same time, limit the electricity bill.

The escalation of energy consumption in wireless networks directly results in the increase of greenhouse gas emission, which has been recognized as a major threat for environmental protection and sustainable development. European Union has acted as a leading flagship in energy saving over the world and targeted to have a 20\% greenhouse gas reduction. China government has also promised to reduce the energy per unit GDP by 20\% and the major pollution by 10\% by the year of 2020. The pressure from social responsibilities serves as another strong driving force for wireless operators to dramatically reduce energy consumption and carbon footprint.
Worldwide actions have been taken. For instance,
Vodafone Group has announced to reduce its CO$_2$ emissions by 50\% against its 2006/7 baseline of 1.23 million tonnes, by the year of 2020\footnote{Information available at: ``http://www.vodafone.com/start/media\_relations/news/group\_press\_releases/2007/01.html''}.


To meet the challenges raised by the high demand of wireless traffic and energy consumption, green evolution has become an urgent need for wireless networks today. As has been pointed out in \cite{TomasEdler:04}, the radio access part of the cellular network is a major energy killer, which accounts for up to more than 70\% of the total energy bill for a number of mobile operators\footnote{The figure is from the energy efficiency solution white paper of Huawei Technologies, ``improving energy efficiency, lower {CO}$_2$ emission and {TCO}'', available at: http://www.huawei.com/green.do.}. Therefore, increasing the energy efficiency of radio networks as a whole can be an effective approach. Vodafone, for example, has foreseen energy efficiency improvement as one of the most important areas that demand innovation for wireless standards beyond LTE \cite{RalfIrmer:09}.

\emph{Green Radio} (GR), a research direction for the evolution of future wireless architectures and techniques towards high energy efficiency, has become an important trend in both academic and industrial worlds. Before GR, there have been efforts devoted to energy saving in wireless networks, such as designing ultra-efficient power amplifier, reducing feeder losses, and introducing passive cooling. However, these efforts are isolated and thus cannot make a global vision of what we can achieve in five or ten years for energy saving. GR, on the other hand, targets at innovative solutions based on top-down architecture and joint design across all system levels and protocol stacks, which cannot be achieved via isolated efforts.

\subsection{Research Activities}
In the academia, several workshops dedicated to green communications have been organized to discuss the future green technologies. For instance, IEEE has two green communication workshops in 2009, in conjunction with ICC'09 and Globecom'09 and at least three more in 2010, in conjunction with ICC'10, PIMRC'10, and Globecom'10, respectively\footnote{ICC, Globecom, and PIMRC are three international conferences under IEEE Communications society, i.e., IEEE International Conference on Communications, IEEE GLOBAL COMMUNICATIONS CONFERENCE
, and IEEE International Symposium on Personal, Indoor and Mobile Radio Communications, respectively.}.

\begin{figure}
\centering
\includegraphics[width = 6in]{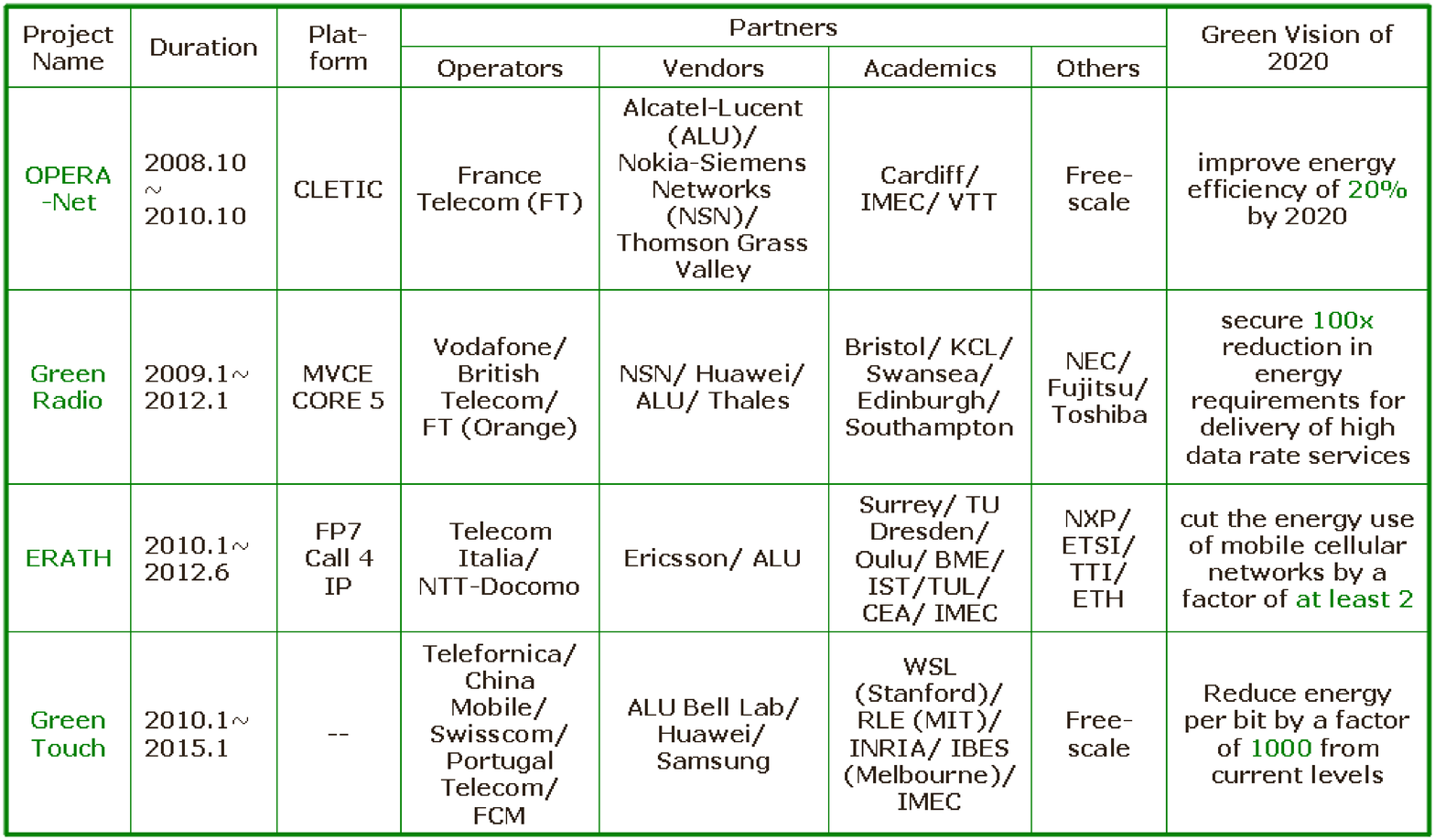}
\caption{International research projects related to green radio.}
\label{fig:projects}
\end{figure}

On the other hand, research projects on GR have sprang up under different international research platforms during the latest years. Fig. \ref{fig:projects} lists some major international projects on GR research.\footnote{Detailed information about these projects can be found at the following addresses: http://www.mobilevce.com/index.htm (MVCE), http://www.greentouch.org (Green Touch), and http://www.opera-net.org/ (OPERA-NET), respectively.} For instance, \emph{Optimizing Power Efficiency in mobile RAdio NETworks} (OPERA-NET), a European research project started in 2008, deals with the energy efficiency in cellular networks. In UK, {GR} is among Core 5 Programs in Mobile VCE since 2009, targeting at parallel evolution of green architectures and techniques. Moreover, \emph{Energy Aware Radio and neTwork tecHnologies} ({EARTH}) \cite{EARTH:09}, one of the integrated projects under European Framework Program 7 Call 4, starts its ball rolling to develop green technologies at the beginning of 2010. Most recently, {GreenTouch}, a consortium of industry, academic, and non-governmental research experts, sets its 5-year research goal to deliver the architecture, specification, and roadmap needed to reduce energy consumption per bit by a factor of 1000 from the current level by the year of 2015.

\subsection{Target of the Article}
\begin{figure}
\centering
\includegraphics[width = 5.5in]{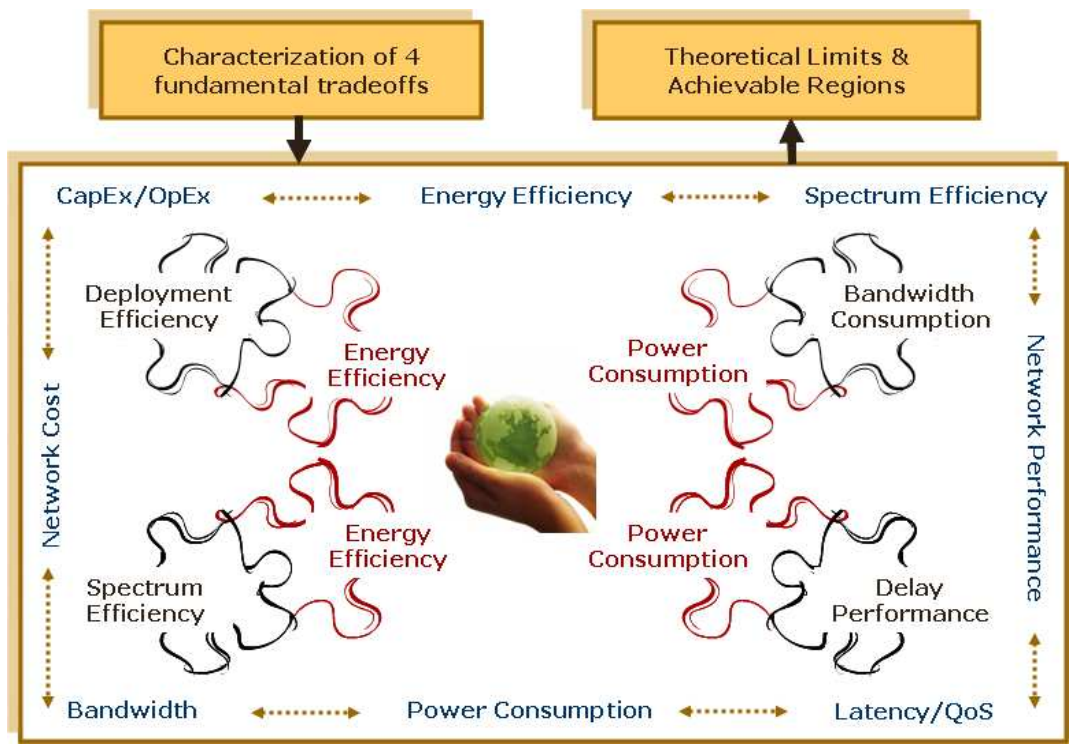}
\caption{Fundamental tradeoffs. }
\label{fig:four_tradeoff_summary}
\end{figure}

GR research is a large and comprehensive area that covers all layers in the protocol stack of wireless access networks as well as the architectures and techniques. Instead of a survey that reaches every aspect of the matter, this article focuses on the fundamental framework for GR research and strings the currently scattered research points using a logical ``rope''.
We propose in this article four fundamental tradeoffs to construct such framework. As depicted in Fig. \ref{fig:four_tradeoff_summary}, they are
\begin{itemize}
\item {\emph{Deployment Efficiency} ({DE}) - \emph{Energy Efficiency} ({EE})} tradeoff: to balance the deployment cost, throughput, and energy consumption, in the network as a whole;
\item {\emph{Spectrum Efficiency} ({SE}) - EE} tradeoff: given a bandwidth available, to balance the achievable rate and the energy consumption of the system;
\item {\emph{Bandwidth} (BW) - \emph{Power} (PW)} tradeoff: given a target transmission rate, to balance the bandwidth utilized and the power needed for the transmission;
\item {\emph{Delay} (DL) - PW} tradeoff: to balance the average end-to-end service delay and the average power consumed in the transmission.
\end{itemize}
By means of the four tradeoffs, key network performance/cost indicators are all stringed together.

\section{Fundamental Framework}
\label{sec:fundamentals}

In this section, we shall elaborate in detail the four tradeoffs that constitute the fundamental framework. As we will see, they actually connect the technologies towards green evolution in different research aspects, such as network planning, resource management, and physical layer transmission scheme design.

\subsection{DE-EE Tradeoff}
DE, a measure of system throughput per unit of deployment cost, is an important network performance indicator for mobile operators. The deployment cost consists of both \emph{capital expenditure} (CapEx) and \emph{operational expenditure} (OpEx). For radio access networks, the CapEx mainly includes infrastructure costs, such as base station equipment, backhaul transmission equipment, site installation, and radio network controller equipment. The key drivers for the OpEx, on the other hand, are electricity bill, site and backhaul lease, and operation and maintenance cost \cite{Johansson:07}. Usually, wireless engineers will estimate the network CapEx and OpEx during network planning. EE, defined as system throughput for unit energy consumption, is mostly considered during network operation.


The two different metrics often lead to opposite design criteria for network planning. For example, in order to save the expenditure on site rental, base station equipment, and maintenance, network planning engineers tend to ``stretch'' the cell coverage as much as possible. However, the path-loss between the base station and mobile users will degrade by $12$ dB whenever the cell radius doubles if the path-loss exponent is four, which induces $12$ dB increase in the transmit power to guarantee the same received signal strength for those users at the cell edges. On the other hand, to provide cellular coverage for a given area, increasing the number of base stations will save the total network transmit power by the same factor. For example, it is shown in \cite{Badic:09} that by shrinking the cell radius from $1,000$ m to $250$ m, the maximum EE of the HSDPA Network will be increased from $0.11$ Mbits/Joule to $1.92$ Mbits/Joule, respectively, corresponding to $17.5$ times of gains. Therefore, to minimize energy radiation, radio resource management engineers favor small cell-size deployment.
From the above discussion, there should be a tradeoff between DE and EE, as shown in Fig. \ref{fig:four_tradeoff_illustration} (a), where each point on the curve corresponds to a cell size, and should be chosen to balance specific DE and EE requirements.

\begin{figure}
\centering
\includegraphics[width = 6.0in]{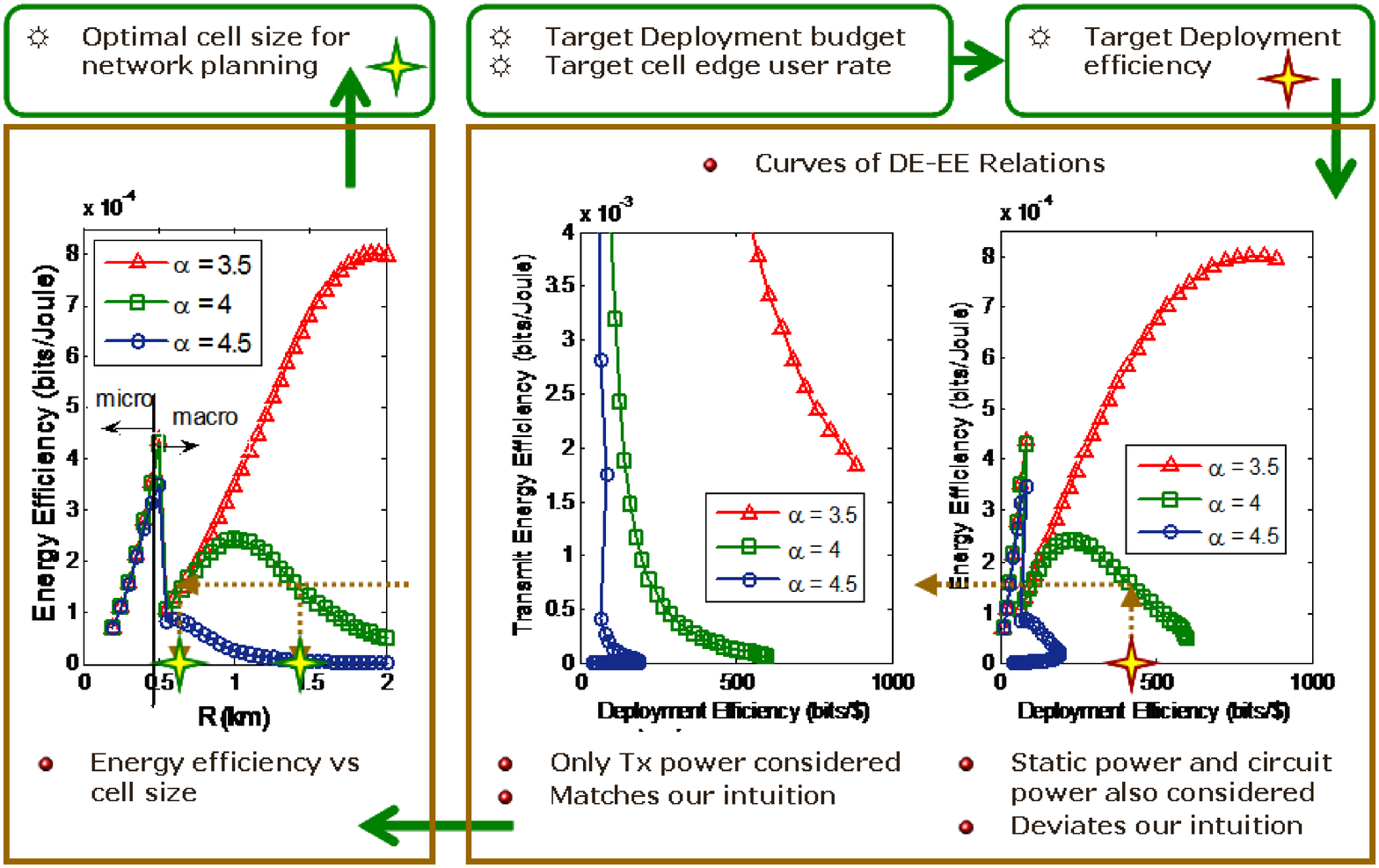}
\caption{Results on DE-EE relation from \cite{GREAT:10-ICC} for different path-loss exponents, $\alpha$. }
\label{fig:DE-EE}
\end{figure}

However, this shape of the curve is correct when only transmission power is considered and the deployment cost scales continuously and proportionally with the cell radius. In reality,
\begin{itemize}
    \item there are limited types of base stations and the equipment cost does not scale proportionally with the target cell size;
    \item the total network energy includes both transmit-dependent energy (e.g. power consumed by radio amplifier) and transmit-independent one (e.g. site cooling power consumption).
\end{itemize}
Therefore, the relation of DE and EE may deviate from the simple tradeoff curve and become more complex when considering practical aspects, as shown in our recent study \cite{GREAT:10-ICC}.
Fig. \ref{fig:DE-EE} summarizes the main result of \cite{GREAT:10-ICC}.
From the right-most plot, there might not always be a tradeoff between DE and EE and the shape of a DE-EE curve depends on the specific deployment scenarios. For the suburb scenario, where the path-loss exponent is small (about $3.5$), the network EE even increases with its DE. For the dense urban scenario, where the path-loss exponent is large (about $4.5$), two different EE values may result in the same DE value, corresponding to very small and very large cell radii, respectively. The former is because of the huge increase in CapEx by increasing the number of sites; the latter is due to the sharply increased electricity bill in OpEx.

Since the shapes of DE-EE curves may not match our intuition, characterizing the curves with practical concerns is helpful to real-world network planning. As shown in Fig. \ref{fig:DE-EE}, for any target network throughput and given deployment budget, we can first calculate the corresponding deployment efficiency, from which we can decide the maximum achievable energy efficiency by looking up the DE value on the DE-EE tradeoff curve; then from the EE versus cell radius curve, we get the corresponding optimal cell size.

No doubt that the current results are still quite preliminary. In the future, research efforts may focus on the following two aspects:
\begin{itemize}
    \item improving the optimal DE-EE frontiers with advanced network architectures;
    \item joint architecture design with advanced transmission schemes and scheduling algorithms to improve the network DE-EE tradeoff relation.
\end{itemize}

For LTE-Advanced or beyond networks, \emph{heterogeneous networks} (HetNet) has been approved as a work item, such as in 3GPP Release 10. With the combination of macro cells and micro/pico/femto cells, the traditionally related functionalities, coverage, and capacity provision can now be decoupled into different tiers of the network. In general, macro cells handle the coverage and mobility issues while micro/pico cells focus on local throughput. It has been shown in \cite{Fettweis:09-VTCfall} that the network EE increases as the density of micro/pico cells grows. On the other hand, the DE aspect of HetNet has been studied in \cite{Johansson:07} for different traffic distributions. From \cite{Johansson:07}, a complementary hot-spot layer of micro/pico cells on top of macro cells has been the most cost-effective architecture for non-uniform spatial traffic. The tradeoff of DE and EE for HetNet, however, is still open.

Another promising candidate for future architectures is \emph{cooperative networks} (CoopNet), where new air-interface techniques, such as relay and \emph{distributed antenna systems} (DAS), are employed. The newly introduced infrastructures, such as relays and remote radio heads, are of much lower cost and smaller coverage compared with macro base stations, which bring mobile users closer to the network and make the deployment more flexible. However, the backhaul cost and signalling overhead may become new killers for energy consumption and system efficiency. Therefore, how much improvement the CoopNet architecture can bring to DE-EE tradeoff needs to be carefully studied.

Moreover, the incorporation of EE oriented user scheduling and radio resource management algorithms on top of HetNet and CoopNet are bound to further improve network utilization efficiency. This is especially important when the spatial traffic distribution is non-uniform and varies with time. Dynamic power control that exploits channel variations has been proved to enhance the link-level power efficiency. Similarly, by extending the idea to network-level, we may introduce dynamic coverage management to exploit traffic variations. Dynamic switch off/on of coverage overlaid cells in low traffic is an example in HetNet while dynamic relay selection or CoMP pattern selection is the counterpart in the CoopNet. As it introduces no extra cost but saves redundant energy consumption, it can improve DE and EE simultaneously. More research efforts on this topic are desired in the future.

\subsection{SE-EE Tradeoff}
SE, defined as the system throughput for unit bandwidth, is a widely accepted criterion for wireless network optimization. The peak value of SE is always among the key performance indicators of 3GPP evolution. For instance, the target downlink SE of 3GPP increases from $0.05$ bps/Hz to $5$ bps/Hz as the system evolves from GSM to LTE. On the contrary, EE, is previously ignored by most of the research efforts and has not been considered by 3GPP as an important performance indicator until very recently. As the green evolution becomes a major trend, energy-efficient transmission becomes more and more important nowadays. Unfortunately, SE and EE are not always consistent and sometimes conflict each other. Therefore, how to balance the two metrics in future systems deserves a careful study.

To characterize the SE-EE tradeoff for point-to-point transmission in \emph{additive white Gaussian noise} (AWGN) channels, Shannon's capacity formula plays the key role. From Shannon's formula, the achievable transmission rate, $R$, under a given transmit power, $P$, and system bandwidth, $W$, is simply $R = W \log_2 (1 + \frac{P}{W N_0})$, where $N_0$ stands for the power spectral density of AWGN. According to their definitions, SE and EE can be expressed as
$\eta_{SE} =\log_2 (1 + \frac{P}{W N_0})$ and $\eta_{EE} ={W \log_2 (1 + \frac{P}{W N_0})}/{(P)}$, respectively. As a result, SE-EE relation can be expressed as
\begin{IEEEeqnarray}{rCl} \label{eqn:SE_EE}
\eta_{EE} = \frac{\eta_{SE}}{(2^{\eta_{SE}} - 1)N_0},
\end{IEEEeqnarray}
which is sketched in Fig. \ref{fig:four_tradeoff_illustration} (a). From the above expression, $\eta_{EE}$ converges to a constant, $1/(N_0\ln 2)$ when $\eta_{SE}$ approaches zero. On the contrary, $\eta_{EE}$ approaches zero when $\eta_{SE}$ tends to infinity.

In practical systems, however, the SE-EE relation is not as simple as the above formula. In particular, circuit power will break the monotonic relation between SE and EE as shown in \cite{Miao_Li:2009,Miao_Li_Koc:2009,GREAT:10-WWRF}. More precisely, if circuit power is considered, the SE-EE curve will turn to a bell shape, as illustrated in Fig. \ref{fig:four_tradeoff_illustration} (c).
From \cite{Miao_Li:2009}, we see that the transmission conditions and strategies, such as the transmission distance, modulation and coding scheme, and resource management algorithms, all have significant impact on the tradeoff of SE and EE.

Nevertheless, the SE-EE relation characterized by equation \eqref{eqn:SE_EE} is only for point-to-point transmission rather than for a network. Further investigation of the energy-efficient transmission policies is expected to obtain more benefit and is crucial for the environmental protection and sustainable development in future wireless cellular systems. Examples of future research topics may include the following aspects
\begin{itemize}
\item characterizing SE-EE tradeoff relation under practical hardware constraints;
\item investigating network SE-EE tradeoff relation in multi-user/multi-cell environments;
\item joint design of physical layer transmission schemes and resource management strategies that will improve the network SE-EE tradeoff relation.
\end{itemize}

The performance limit predicted by theoretical analysis may not be achieved in real systems due to the practical hardware constraints. For instance, the typical energy conversion efficiency\footnote{Also known as drain efficiency, defined as the ratio of output power over input power.} of a power amplifier in current base stations is less than $40$\%. Moreover, the limited linearity regions of power amplifiers also set a constraint on the transmitted signals, such as the peak-to-average power ratio. How these issues would affect the SE-EE tradeoff is not clear yet.
Therefore, a more detailed modeling of the equipment level energy consumption and practical constraints in hardware devices and transmission signals will help us to find practically achievable SE-EE regions. The gaps between the theoretical limits and achievable regions may further guide the design of future wireless networks.

For the multi-user/multi-cell cases, inter-user interference or inter-cell interference may break the fundamental assumptions in the point-to-point cases. An interesting extension of SE-EE tradeoff relation to multi-cell scenarios with inter-cell interference has been studied in \cite{Miao_Li_Koc:2009}. From \cite{Miao_Li_Koc:2009}, the interference power generated by the neighboring cells not only reduces the maximum achievable EE but also degrades SE and EE. As we can imagine, the higher the interference level, the larger the degradation would be. In this case, the results from the simple point-to-point case are not applicable and a systematic approach towards the multi-user/multi-cell systems shall be developed to build the theoretical fundamentals of the energy-efficient wireless transmissions.

Energy efficient transmission, from the point of view of resource management, can be interpreted as assigning the \emph{right} resource to transmit to the \emph{right} user at the \emph{right} time. Cross-layer optimization techniques, which have been proved useful, may also help to design resource allocation or user scheduling algorithms that optimize the achievable SE-EE tradeoff. A comprehensive survey on the techniques for energy-efficient wireless communication from time, frequency, and spatial domains can be found in \cite{Miao_Li:2009} and it may serve as a good tutorial. In advanced network architectures, such as HetNet and CoopNet, the system may benefit even more from the joint design of physical transmission and resource management. Our recent work in \cite{GREAT:10-WWRF} presents initial results in relay-assisted cooperative systems.

\subsection{BW-PW Tradeoff}
BW and PW are the most important but limited resources in wireless communications. From the Shannon's capacity formula, the relation between the transmit power and the signal bandwidth for a given transmission rate, $R$, can be expressed as
\begin{IEEEeqnarray}{rCl}
P = W N_0 (2^{\frac{R}{W}} - 1).
\end{IEEEeqnarray}
The above expression shows a monotonic relation between PW and BW as sketched in Fig. \ref{fig:four_tradeoff_illustration} (b). It can be easily seen from the above expression that the minimum power consumption is as small as $N_0 R \ln 2$ if there is no bandwidth limit.

The fundamental BW-PW relation in Fig. \ref{fig:four_tradeoff_illustration} (b) shows that, for a given data transmission rate, the expansion of the signal bandwidth is preferred in order to reduce the transmit power and thus achieves better energy efficiency. In fact, the evolution of wireless systems exhibits the same trend for bandwidth demand. For example, in GSM systems, bandwidth per carrier is $200$ kHz while it is $5$ MHz in UMTS systems. In future wireless systems, such as LTE or LTE-Advanced, system bandwidth is $20$ MHz and may even reach up to as wider as $100$ MHz if some techniques, such as \emph{carrier aggregation} (CA)\footnote{Carrier aggregation (CA) is a technique that enables aggregation of multiple component carriers (basic frequency blocks) into overall wider bandwidth. CA is among the main features in LTE-Advanced.}, are used.

\begin{figure}
\centering
\includegraphics[width = 5.5in]{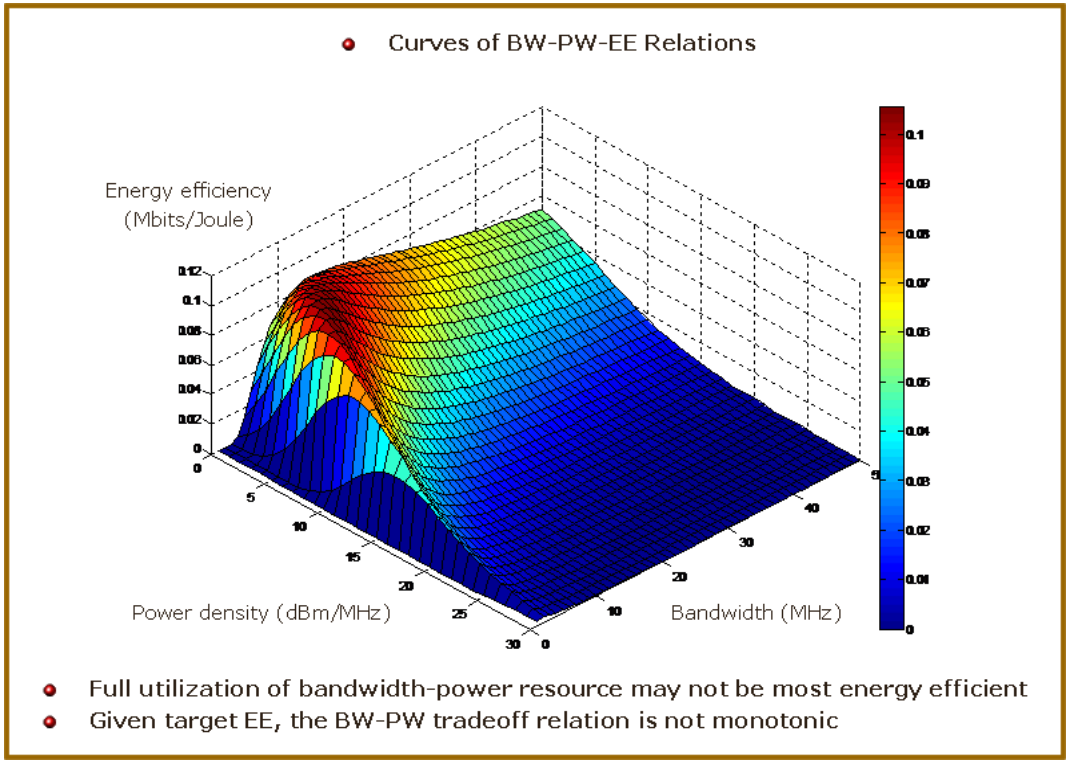}
\caption{Results on BW-PW-EE relation for fixed transmission rate.}
\label{fig:BW-PW}
\end{figure}

The BW-PW relation is also crucial to radio resource management. In \cite{Grace:09-CROWNCOM}, it has been exploited to determine the ``green'' transmission strategy, which first senses and aggregates the unused spectrum using \emph{cognitive radio} (CR) techniques, and then adjusts the modulation order according to the available BW each time. However, in practical systems, the circuit power consumption, such as filter loss, actually scales with the system BW, which entangles the BW and PW relation as shown in Fig. \ref{fig:four_tradeoff_illustration} (d). Furthermore, Fig. \ref{fig:BW-PW} illustrates a visual example of the 3-dimension relation among PW, BW, and EE. From the figure, we have the following two observations.
\begin{itemize}
    \item If the circuit PW scales with the transmission BW (fixed power spectrum density), fully utilization of the bandwidth-power resources may not be the most energy-efficient way to provide the wireless transmission under fixed transmission rate.
    \item Given a target EE, the BW-PW relation is non-monotonic.
\end{itemize}

Although the BW-PW tradeoff has been noticed decades ago, there are still many opening issues that deserve future investigation. Some of them are
\begin{itemize}
    \item advanced techniques for BW-PW tradeoff with practical concerns;
    \item novel network architectures and algorithms to improve BW-PW tradeoff.
\end{itemize}

As we know, the 2G and 3G wireless communication systems, such as GSM and UMTS, use fixed BW transmission, leaving no space for dynamic BW adjustment. With the evolution of wireless technologies, the future deployment of LTE or LTE-Advanced systems provides more flexibility on the spectrum usage so that the transmission BW can be tuned for different applications. Meanwhile, technologies, such as spectrum re-farming\footnote{Spectrum re-farming is more like a government action to support more efficient use of wireless spectrum via reassigning 2G spectrum to 3G applications. For instance, it is now possible to deploy UMTS (3G system) on $900$ MHz (2G spectrum).}, CA, and \emph{software defined radio} (SDR) based CR techniques, are maturing to support the flexible use of BW. However, the implementation and integration of these technologies will incur extra overhead in practical systems. For example, CA requires multiple \emph{radio frequency} (RF) chains and CR needs additional energy for sensing. Therefore, we shall pay more attention to how these technologies can be integrated efficiently.

On the other hand, the deployment of advanced network architecture may also change the shape of the BW-PW tradeoff frontier. In particular, the deployment of CoopNet and HetNet introduces additional infrastructure nodes into the network; consequently, the BW and PW planning will be different from the conventional network architectures. Hence, the BW-PW tradeoff with advanced resource management algorithms under new network architectures deserves future research. In addition, with the combination of CA and CR techniques, cross-layer approaches that jointly consider dynamic BW acquisition and BW-PW tradeoff will certainly play important roles in the future design.

\subsection{DL-PW Tradeoff}
In the tradeoffs described above, the metrics such as DE, SE, and BW, are either system efficiency or resource, which are more physical layer oriented. Different from these metrics, DL, also known as service latency, is a measure of QoS and user experience and is closely related to the upper layer traffic types and statistics. As a result, the design of transmission schemes shall cope with both channel and traffic uncertainties, which makes the characterization of DL-PW tradeoff more complicated.

In early mobile communication systems, such as GSM, the service type is very limited and focuses mainly on voice communications. The traffic generated in voice service is continuous and constant where fixed rate coding and modulation schemes are good enough. In this case, the DL between the transmitter and the receiver mainly consists of signal processing time and propagation delay. Hence, there is not much we need to do. However, the types of wireless services become diverse as technologies evolve and the ability of mobile terminals enhances the popularity of mobile http service, multimedia message service, and multimedia video service. The future networks must be with various applications and heterogeneous DL requirements. Therefore, in order to build a green radio, it is important to know when and how to trade tolerable DL for low power.

To understand the DL-PW tradeoff, let us start with the simplest case first excluding the impact of both channel and traffic dynamics. For point-to-point transmission over AWGN channels, Shannon's formula tells us that $R = W \log_2 (1 + \frac{P}{W N_0})$ bit information are transmitted each second; hence, it takes $t_b=1/R$ second to transmit a bit. Therefore, the average power per bit can be expressed as
\begin{IEEEeqnarray}{rCl}
P_b = W N_0 t_b \left( 2^{\frac{1}{t_b W}} - 1 \right).
\label{eqn:DL-PW}
\end{IEEEeqnarray}
The above expression shows a monotonically decreasing relation between per bit PW and DL as sketched in Fig. \ref{fig:four_tradeoff_illustration} (b). Also note that $\frac{1}{t_bW}=\frac{R}{W}$ can be regarded as modulation level for an uncoded communication system. Then the transmit power per bit decreases as the modulation level reduces. However, as in all other three tradeoff relations, once we take practical concerns into consideration, such as circuit power, the tradeoff relation usually deviates from the simple monotonic curve and it may appear like a cup shape as sketched in Fig. \ref{fig:four_tradeoff_illustration} (d).

DL-PW relation with traffic dynamics is more complicated. In this case, the service DL should include both the waiting time in the traffic queue and the time for transmission, the sum of two part is also known as queueing DL. In addition, when traffic flow is considered, average DL per packet will be used instead of average DL per bit. The basic tradeoff in \eqref{eqn:DL-PW} has been extended to the finite packets scheduling in \cite{UB_Gamal:2002}. A lazy schedule was proposed to minimize the total transmission power while guaranteeing the transmission of all packets to be finished before a pre-determined time. A benchmark paper \cite{Berry_Gallager:2002} takes both channel uncertainties and random traffic into consideration. However, the mathematical model there is very complicated since both information theory and queueing theory are involved. Nevertheless, the results there is only for point-to-point case\footnote{There are progresses on the DL-PW tradeoff research in recent years. Interested readers may refer to the following link: ``http://ee.usc.edu/stochastic-nets/wiki/dokuwiki-2008-05-05/doku.php''.} and more open issues need to be addressed, including
\begin{itemize}
    \item DL-PW tradeoff for heterogeneous DL requirements in multi-user/multi-cell scenarios;
    \item joint design of physical layer transmission schemes and resource management to improve DL-PW tradeoff with consideration of practical concerns;
    \item simplified and insightful but approximate mathematical models for DL-PW relation.
\end{itemize}

From queueing theory, we know that the average DL of a packet queue is determined by the statistics of the traffic arrivals and departures. Usually, the departure rates are closely related to the transmission schemes and the radio resources available. In multi-user/multi-cell environments, however, the system resources are shared among different users and also among various application streams, which makes the departure rates of different queues correlated with each other. Consequently, network DL-PW relation needs to be considered and the mathematical model becomes even more complicated. In general, there is no closed-form expression available to show the direct relation between DL and PW. Therefore, the investigation of simplified but approximate models is desired to provide insights for practical system design. On the other hand, due to correlation among queues, user scheduling and resource allocation algorithms are crucial to control the operation point that maximizes network power efficiency while balancing the heterogeneous DL requirements.

\section{Conclusions}
\label{sec:con}
\begin{figure}
\centering
\includegraphics[width = 5.5in]{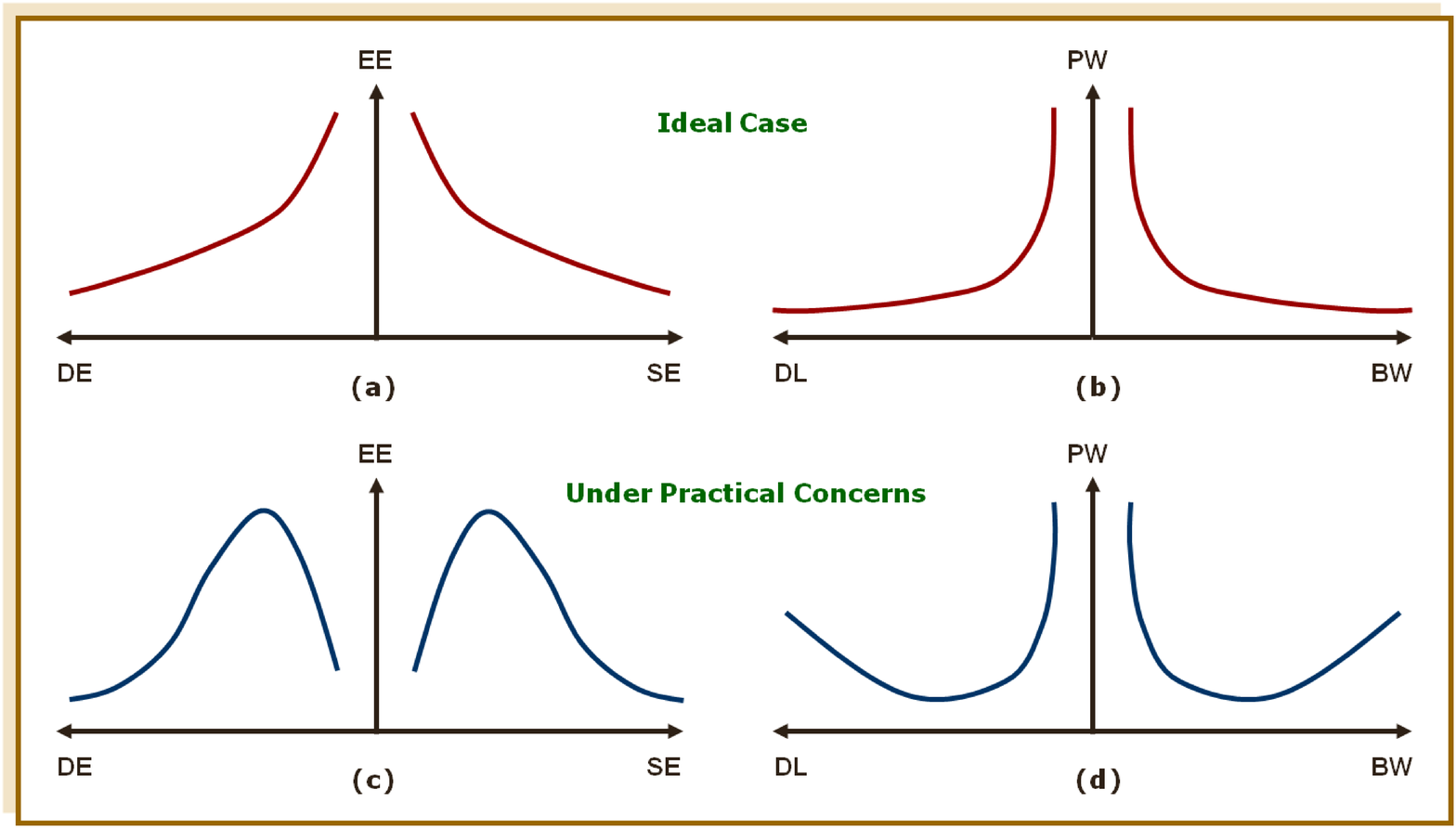}
\caption{Sketch of the four tradeoff relations without and with practical concerns. }
\label{fig:four_tradeoff_illustration}
\end{figure}

\begin{figure}
\centering
\includegraphics[width = 6.2in]{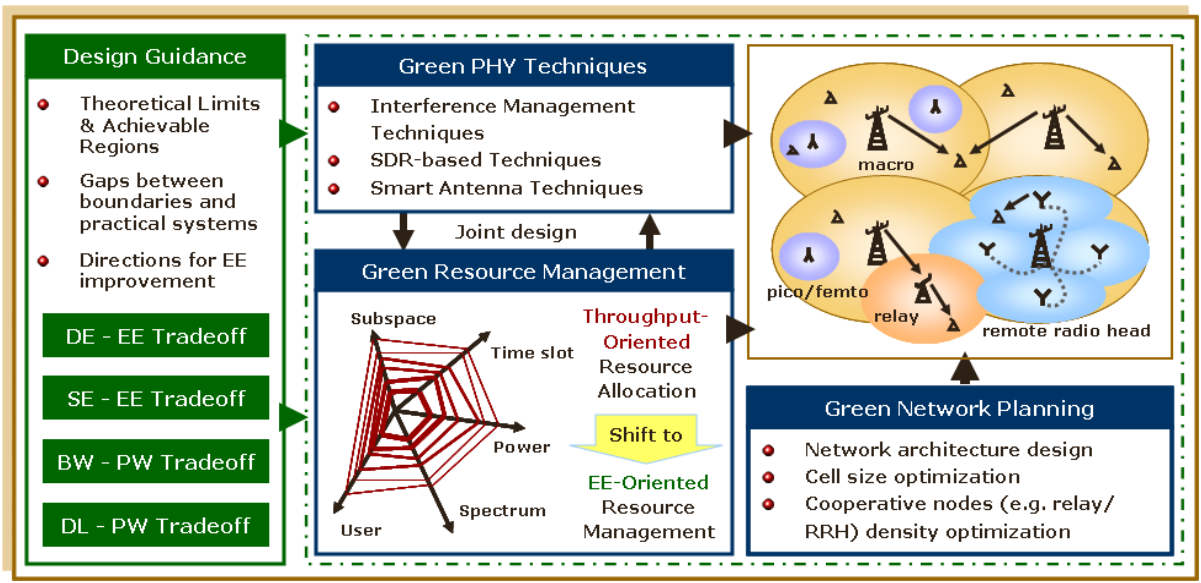}
\caption{An overview of how the fundamental framework guides specific system designs. }
\label{fig:whole_pic}
\end{figure}

In this article, we have proposed a framework for GR research to integrate the fundamental connections that are currently scattered. Four fundamental tradeoffs constitute the skeleton of the framework.
We have shown that, in practical systems, the tradeoff relations usually deviate from the simple monotonic curves derived from Shannon's formula as summarized in Fig. \ref{fig:four_tradeoff_illustration}. Moreover, most of the existing literature mainly focuses on the point-to-point single cell case. Therefore, the tradeoff relations under more realistic and complex network scenarios deserve future investigation. The insights, such as how to improve the tradeoff curves as a whole and how to tune the operation point on the curve to balance the specific system requirements, are expected to guide the practical system designs towards green evolution, which will be our next steps following this piece of work.
Fig. \ref{fig:whole_pic} demonstrates a whole picture of how the proposed framework will impact the green design of future systems.

As the market develops, wireless networks will continue to expand in the future. Green evolution, as a result, will continue to be an urgent demand and inevitable trend for operators, equipment manufacturers, as well as other related industries.
Progresses in fundamental GR research, as outlined in this article, will certainly help in making a green future.


\newpage

\bibliographystyle{IEEEtran}
\bibliography{IEEEabrv,mybib_commag}

\newpage

\newpage


\newpage

\newpage

\newpage

\newpage

\end{document}